\documentclass[twocolumn,preprintnumbers,amsmath,amssymb]{revtex4}

\usepackage{graphicx}
\usepackage{dcolumn}
\usepackage{bm}
\usepackage{here}
\usepackage{float}
\begin{document}
\title{General Connectivity Distribution Functions for Growing Networks\\
with Preferential Attachment of Fractional Power}

\author{
Kazumoto Iguchi
and 
Hiroaki S. Yamada\cite{byline1}}

\address{{\it KIRL, 70-3 Shinhari, Hari, Anan, Tokushima 774-0003, Japan\/}}
\address{{\it YPRL, 5-7-14-205 Aoyama, Niigata 950-2002, Japan\/}\cite{byline1}}

\date{\today}

\begin{abstract}
We study the general connectivity distribution functions for growing networks
with preferential attachment(PA) of fractional power, $\Pi_{i} \propto k^{\alpha}$, using Simon's method. 
We first show that the heart of the previously known methods of the rate equations for
the connectivity distribution functions is nothing but Simon's method for
word problem.
Secondly, we show that for the case of fractional $\alpha$ the $Z$-transformation of the
rate equation provides a fractional differential equation of new type,
which coincides with that for PA with linear power, when $\alpha = 1$.
We show that to solve such a fractional differential equation we need define
a transidental function $\Upsilon (a,b,c;z)$ that we call {\it upsilon function}.
Most of all the previously known results are obtained consistently in the frame work of a unified theory.
\end{abstract}

\pacs{02.50.Cw, 05.40.-a, 05.50.+q, 87.18.Sn}
\maketitle



\section{Introduction}
Random network models have been a prototype model
for a complex network system for a long time\cite{ErdosRenyi}.
However, at nearly the end of 1990's  
the scale-free networks were discovered
from studying the growth of the internet geometry and topology.
After the discovery, scientists have found that
many real systems such as 
internet topology,
human sexual relationship, 
scientific collaboration,
economical network, and so on
show the scale-free network topology.
\cite{Strogatz01,Barabasi02,AlbertBarabasi02,
BaraBona03,Newman03,DorogovMendes03,Boccaletti03}.
The most characteristic property of the scale-free network
is the scaling property that
the distribution function $P(k)$ of numbers of links with degree $k$ is given by
$P(k) \propto k^{-\gamma}$
where $2< \gamma < \infty$.
To describe the scaling law in the networks
there have appeared many studies, using various statistical and mathematical methods 
\cite{BarabasiAlbert99,DorogovMendes00,Krapivsky00,Dorogovtsev00,
BarabasiAlbert00,Cohen00,Callaway01,BornEbel01,
BianconiBara01,Goh01,Liu02,Milo02,Przytycka04,Song05,
Catanzaro05,White05,TakemotoOosawa05,Moore06,Bu07}.

In those studies 
Barab\'{a}si and Albert\cite{BarabasiAlbert99} first introduced a growing network model 
(called the BA-model).
They introduced the concept of growing networks with preferential attachment (PA) 
that is proportional to link degree $k$ and the system 
exhibits a scaling property of $P(k) \propto k^{-3}$.
In terms of time development of the connectivity distribution function,
Krapivsky, Redner and Leyvraz\cite{Krapivsky00}
and Dorogovtsev and Mendes\cite{Dorogovtsev00}
have generalized the PA to include
the more general PA that is proportional to $k^{\alpha}$ 
and to that is proportional to attractiveness $A$ in addition 
to linear PA such as $A + k$,
respectively.
Liu, Lai and Ye\cite{Liu02} 
have given a most direct generalization of the BA-model
to include the $k^{\alpha}$-PA in the spirit of 
Krapivsky, Redner and Leyvraz\cite{Krapivsky00}. 

On the other hand, to show the scale-free nature and to obtain
the scaling exponent of the distribution function,
Bornholdt and Ebel\cite{BornEbel01} have presented another
method based on the famous Simon's method\cite{Simon55}.
They obtained the exponent $\gamma = 1 + 1/(1-\beta)$,
where $\beta$ ($0< \beta < 1$) is the probability that
a newly attaching node is exerted into the system,
while $1-\beta$ is the probability that a new link is
attached between the existing nodes.
Therefore, if $\beta = 1$ then the system must coincide with
the one studied by Krapivsky, Redner and Leyvraz\cite{Krapivsky00}.
However, it does not do so,
because in this limit the former provides $\gamma = \infty$,
while the latter provides $\gamma = 3$.
Hence, there is something wrong in their study.

Their flaw lies in the two assumptions that were borrowed from
those of Simon for the words problem in which
the probability that the adding word is a new word is given $\beta$
while the probability that the adding word is the word that already exists $i$-times
is given $1-\beta$\cite{Simon55}.
In the word problem, there is only a single PA of word to the system
and therefore, there is no difference in the sense of attachment 
of the new word and existing words since both are words.
On the other hand, when we apply the method to 
a network model, we have to be careful 
because in the network model, there are both attachments of nodes and links.
We should not confuse with such attachments so that
a node is exerted into the system at each time step and
at the same time links are provided accordingly.
We cannot separate both procedures using a single
parameter $\beta$ so that
the attachment of a node is given by $\beta$ and
the attachment of links by $1-\beta$.
This is unacceptable.
However, the mathematical ideas that they adopted is very
interesting in its own right\cite{Footnote1}.

In this paper, we would like to study 
the growing network models with PA with fractional power,
using Simon's method\cite{Simon55}
in order to unify most of the previously known methods
and results studied in the literature
\cite{BarabasiAlbert99,DorogovMendes00,Krapivsky00,Dorogovtsev00,
BarabasiAlbert00,Cohen00,Callaway01,BornEbel01,
BianconiBara01,Goh01,Liu02,Milo02,Przytycka04,
Song05,Catanzaro05,White05,TakemotoOosawa05,Moore06,Bu07} 
and to give new results.
We would like to shed a new light on the problem.

The organization of the paper is the following.
In Sec.II, we review the growing network models with various types of PA.
In Sec.III, we introduce the rate equations for the connectivity
distribution functions for the entire network and a particular node, respectively.
In Sec.IV, we solve the rate equation for the connectivity
distribution function for the entire network.
In Sec.V, we solve the rate equation for the connectivity
distribution function for $i$th node.
In Sec.VI, a conclusion is made.


\section{The evolving network models}
Random network models of Erd\"{o}s and R\'{e}nyi\cite{ErdosRenyi}
have been standard models for complex network systems,
until evolving network models have been found by 
Barab\'{a}si and Albert\cite{BarabasiAlbert99}.
In the former, the size of the network $N$ is fixed and
the links are distributed with probability $p$ such that
each link has the mean connectivity $\langle k \rangle = pN$.

In the latter, the network grows from a set of $m_{0}$ seed nodes,
putting a newly added node from which $m$ new links are distributed
to the existing nodes. The way of adding a link depends on
the PA of 
$$\Pi_{i} = \frac{k_{i}}{\sum_{j=1}^{N-1}k_{j}}.  \eqno{(1)}$$
This yields an evolution equation for the link number $k_{i}$ of the $i$th node:
$$\frac{dk_{i}}{dt} = m\Pi_{i} = \frac{m k_{i}}{\sum_{j=1}^{N-1}k_{j}}
=  \frac{k_{i}}{2t},  \eqno{(2)}$$
where we have a trivial relation $\sum_{j=1}^{N-1}k_{j} = 2mt - m$.

Recently, Liu, Lai, and Ye\cite{Liu02} have generalized the above PA
to the following:
$$\Pi_{i} = \frac{k_{i}^{\alpha}}{\sum_{j=1}^{N-1}k_{j}^{\alpha}},  \eqno{(3)}$$
and the evolution equation to the following:
$$\frac{dk_{i}}{dt} = \bar{m}\Pi_{i} 
= \frac{\bar{m} k_{i}^{\alpha}}{\sum_{j=1}^{N-1}k_{j}^{\alpha}}
=  \bar{m}\frac{k_{i}^{\alpha}}{\mu_{\alpha} t},  \eqno{(4)}$$
where $0 < \alpha < 1$, $\bar{m}$ is the average number of attached links 
and we have used the relation
$$\sum_{j=1}^{N}k_{j}^{\alpha} \equiv \mu_{\alpha} t,  \eqno{(5)}$$
where $1 < \mu_{\alpha}< 2\bar{m}$ for $0 < \alpha < 1$
and $\mu_{1} = \langle k \rangle =2\bar{m}$ for $\alpha = 1$.
The solution of the above evolution equation has been given by them.

This formalism is basically the same as that of 
Krapivsky, Redner and Leyvraz\cite{Krapivsky00}
as discussed by Liu, Lai, and Ye\cite{Liu02} in their appendix,
since the former solved the time evolution of Eq.(4)
while the latter solved the rate equation of the
connectivity distribution function which will be discussed
in the next section.
Furthermore, it is also almost equivalent to the
formalism of 
Dorogovtsev, Mendes and Samukhin\cite{Dorogovtsev00}
and 
Bianconi and Barab\'{a}si\cite{BianconiBara01}.

To be more specific to this point,
the PA for the model of Dorogovtsev, Mendes and Samukhin\cite{Dorogovtsev00}
is defined as follows:
$$\Pi_{i} = \frac{A + k_{i}}{\sum_{j=1}^{N-1}(A + k_{j})},  \eqno{(6)}$$
where $A$ is called the initial {\it attractiveness} of node $i$,
which describes the ability of the $i$th node to receive links from
other nodes, i.e., the probability for young nodes to get new links. 
On the other hand, 
the PA for the model of Bianconi and Barab\'{a}si\cite{BianconiBara01} is given by
$$\Pi_{i} =  \frac{\eta_{i} k_{i}}{\sum_{j=1}^{N-1}\eta_{j} k_{j}},  \eqno{(7)}$$
where $\eta_{i}$ is called the {\it fitness} of node $i$,
which describes the discrimination of different types of nodes.
Although in the original Bianconi-Barab\'{a}si model\cite{BianconiBara01},
the fitness is distributed at random with a distribution of $\rho(\eta)$,
we restrict ourselves to the case of the fixed fitness for each node.
In this sense, we would like to call such fixed value of fitness the {\it preference}
throughout this paper.
Thus, we are able to recognize that the concepts
of attractiveness and preference are essentially equivalent to each other,
since one can expand the preference as $\eta_{i} = 1 + A/k_{i}$. 

In this way, we are led to define the most general form of the PA
for a finite system of $N$ nodes:
$$\Pi_{i} =  \frac{\eta_{i} k_{i}^{\alpha}}{\sum_{j=1}^{N-1}\eta_{j} k_{j}^{\alpha}},  \eqno{(8)}$$
where $\alpha$ can be any real number of $-\infty < \alpha < \infty$ in general.


\section{The Rate Equation}
\subsection{Definition of the Rate Equation}
Let us first define that at each instant of time a node is added to the system
such that the total number of nodes is given by
$N(t) = t$.
Let us denote by $N_{k}(i,t)$ the distribution of the connectivity (= degree)
$k$ at the node $i$ and define $N_{k}(i,i) = \delta_{k,m}$,
which means that the node $i$ starts with $k_{i}(0)=m$ when it is born\cite{Footnote2}.

Following both Krapivsky, Redner and Leyvraz\cite{Krapivsky00}
and 
Dorogovtsev, Mendes and Samukhin\cite{Dorogovtsev00},
we can define the rate (or master) equation for the time evolution
of the connectivity distribution function as follows.
Let us define the probability that the node $i$ 
receives exactly $l$ new links of the $m$ injected as
$$p_{k,l} = 
\left( \begin{array}{c}
m \\ l
\end{array} \right)
(\Pi_{k})^{l}(1-\Pi_{k})^{m-l}, \eqno{(9)}$$
where $\Pi_{k}$ is the PA given in the previous section,
regarding $k_{i}$ as $k$.
Then the rate equation is given by
$$N_{k}(i,t+1)  = \sum_{l=0}^{m}p_{k,l}N_{k-l}(i,t). \eqno{(10)}$$

We would like to note here that Eq.(9) describes that at each time step, each node
can accept $l$ new links from a same single node, in general.
However, in the most of models the number  of such receiving links is $0$ or $1$.
Therefore, at this moment the above setting seems somewhat strange
but the selection of the receiving link number can be realized from
expansion of the PA [see Eq.(13)].

\subsection{The Rate Equation for the Entire Network}
Let us define the connectivity distribution of the entire network by
$$N_{k}(t) =\frac{1}{t} \sum_{i=1}^{t}N_{k}(i,t).   \eqno{(11)}$$
Summing up Eq.(10) over $i$ from $1$ to $t$,
we obtain
$$(t+1)N_{k}(t+1) -tN_{k}(t+1,t+1) = \sum_{l=0}^{m} t p_{k,l} N_{k-l}(t). \eqno{(12)}$$
Now we expand the right hand side with respect to PA $\Pi_{k}$ as
$$rhs = t(1-\Pi_{k})^{m}N_{k}(t) + m t \Pi_{k-1}(1-\Pi_{k})^{m-1}N_{k-1}(t) + \cdots$$ 
$$= t(1-m\Pi_{k})N_{k}(t) + mt \Pi_{k-1}N_{k-1}(t) + \cdots$$
$$= \left(t-\frac{m\eta_{k} k^{\alpha}}{\mu_{\alpha}}\right)N_{k}(t) 
+ \frac{m\eta_{k-1} (k-1)^{\alpha}}{\mu_{\alpha}}N_{k-1}(t) + \cdots,   \eqno{(13)}$$
where we have defined 
$$\sum_{i=1}^{t} \eta_{i} k_{i}^{\alpha}
= \sum_{k=1}^{\infty}\eta_{k} k^{\alpha} N_{k}(t) \equiv \mu_{\alpha} t. \eqno{(14)}$$
It is equivalent to
$$\frac{1}{t}\sum_{i=1}^{t} \eta_{i} k_{i}^{\alpha} 
= \sum_{k=1}^{\infty}\eta_{k} k^{\alpha} n_{k} = \mu_{\alpha}, \eqno{(15)}$$
where $1< \mu_{\alpha} <2m\langle \eta_{k}\rangle$ for $0 < \alpha <1$
and the appearance of the distribution function $n_{k}$ in the second expression 
is due to the definition that $n_{k} = \frac{1}{t}\sum_{i=1}^{t}\delta_{k,k_{i}}$.
If $\eta_{k} = 1$, then by definition $\mu_{\alpha} =\langle k^{\alpha} \rangle$. 
The verification of Eq.(15) will be discussed in the next subsection[see Eq.(30)].
Here, we would like to note that the $\mu_{\alpha}$ in Eqs.(14) and (15) is different 
from the one in Eq.(5), since in the latter there is the preference factor in the expression.

From Eqs.(12) and (13), we obtain
$$(t+1)N_{k}(t+1) - t\delta_{k,m} 
= \left[t-\frac{m\eta_{k} k^{\alpha}}{\mu_{\alpha}}\right] N_{k}(t) $$
$$+ \frac{m\eta_{k-1} (k-1)^{\alpha}}{\mu_{\alpha}} N_{k-1}(t), \eqno{(16)}$$
where higher terms are omitted.
Exchanging some terms in both sides, we have
$$(t+1)N_{k}(t+1) - t N_{k}(t) $$
$$= \frac{m}{\mu_{\alpha}}[ -\eta_{k}k^{\alpha} N_{k}(t) 
+ \eta_{k-1}(k-1)^{\alpha} N_{k-1}(t)] + t\delta_{k,m}. \eqno{(17)}$$
This is the most general form for the rate equation for our
purpose here.
In these equations, $N_{k}(t)$ are defined for $k \ge m$.

Now, supposing that the time is as sufficiently large as 
$t >> 1$ such that $t+1 \approx t$, 
the left hand side becomes $t[N_{k}(t+1) -  N_{k}(t)]$.
Then dividing both sides by $t$, we obtain
$$N_{k}(t+1) - N_{k}(t) $$
$$= \frac{m}{\mu_{\alpha}t}[ -\eta_{k}k^{\alpha} N_{k}(t) 
+ \eta_{k-1}(k-1)^{\alpha} N_{k-1}(t)] + \delta_{k,m}. \eqno{(18)}$$
This is exactly the same form of the evolution equation that
was studied by Simon\cite{Simon55,BornEbel01},
apart from the coefficient of the last term in the right hand side;
in the Simon model, the last term is $\beta \delta_{k,0}$, where 
$\beta$ is the probability that the newly adding word is a new word
while $1-\beta$ the probability that it is one of already existing words.

\subsection{The Rate Equation for the Connectivity Distribution Function}
We follow the same argument for derivation of the 
rate equation for the connectivity distribution function, $N_{k}(i,t)$.
Let us expand the right hand side of Eq.(10) with respect to $l$.
We obtain
$$N_{k}(i,t+1)  = p_{k,0}N_{k}(i,t) + p_{k,1}N_{k-1}(i,t) + \cdots.\eqno{(19)}$$
Expanding the $p_{k,0}$ and $p_{k,1}$ with respect to the $\Pi_{k}$
in the same way, we obtain
$$N_{k}(i,t+1)  = \left[1 - \frac{m\eta_{k}k^{\alpha}}{\mu_{\alpha} t}\right]N_{k}(i,t) $$
$$+ \frac{m\eta_{k-1}(k-1)^{\alpha}}{\mu_{\alpha} t}N_{k-1}(i,t) + \cdots.\eqno{(20)}$$

This is the most general generalization for the rate equation 
of the connectivity distribution functions.
It includes all the models studied in the previous literature
\cite{BarabasiAlbert99,DorogovMendes00,Krapivsky00,Dorogovtsev00,
BarabasiAlbert00,Cohen00,Callaway01,BornEbel01,
BianconiBara01,Goh01,Liu02,Milo02,Przytycka04,
Song05,Catanzaro05,White05,TakemotoOosawa05,Moore06,Bu07}.


\section{The Solution of the Rate Equations for the Entire Network}

\subsection{Simon's method}
Let us now consider the steady state solution of the rate equation
for the entire network of Eq.(17).
Let us first note that if we expand the rate equation
with respect to time $t$, then we are able to obtain
the continuous rate equation such as those studied by
Krapivsky, Redner and Leyvraz\cite{Krapivsky00}
and 
Dorogovtsev, Mendes and Samukhin\cite{Dorogovtsev00}. 
Therefore, we can use their methods here.
However, as is mentioned before,
the rate equation has exactly the same form that was first studied by Simon
for the words problem\cite{Simon55}.
Therefore, we can apply his mathematical method to this problem as well.

Following Simon\cite{Simon55}, 
for the connectivity distribution function for the steady state
we can invoke the following relation:
$$\frac{N_{k}(t+1)}{N_{k}(t)} = \frac{t+1}{t}.   \eqno{(21)}$$
This mean that the $N_{k}(t)$ grows like $N_{k}(t) \propto t$.
Next, let us define
$$\frac{N_{k}(t+1)}{N_{k-1}(t+1)} = \frac{N_{k}(t)}{N_{k-1}(t)} \equiv \beta_{k}.   \eqno{(22)}$$
Substituting the relation of Eq.(21) into the left hand side of Eq.(18)
and the relation of Eq.(23) into the right hand side of Eq.(18),
we obtain
$$\left[\frac{t+1}{t}-1\right]N_{k}(t)  = \frac{1}{t}N_{k}(t) $$
$$= \frac{m}{\mu_{\alpha}t}\left[\frac{\eta_{k-1}(k-1)^{\alpha}}{\beta_{k}}
-\eta_{k}k^{\alpha}\right] N_{k}(t) + \delta_{k,m}.                   \eqno{(23)}$$
Hence, we can define the steady state solution $n^{*}_{k}$ that is independent of $t$
such as
$$N_{k}(t) \equiv n^{*}_{k}t.   \eqno{(24)}$$

For $k>m$, we can derive the following for the steady state solution $n^{*}_{k}$:
$$\beta_{k} = \frac{\eta_{k-1}(k-1)^{\alpha}}{\frac{\mu_{\alpha}}{m}+\eta_{k}k^{\alpha}}
= \frac{n^{*}_{k}}{n^{*}_{k-1}}. \eqno{(25)}$$
For $k=m$, since $N_{k}(t) = 0$ for $0 < k < m$, Eq.(18) yields
$$\frac{1}{t} N_{m}(t) = 1 -\frac{\eta_{m}m^{1+\alpha}}{\mu_{\alpha}t} N_{m}(t). \eqno{(26)}$$
Therefore, we have
$$N_{m}(t) = \frac{\frac{\mu_{\alpha}}{m}}{\frac{\mu_{\alpha}}{m}+m^{\alpha}\eta_{m}} t \equiv n^{*}_{m} t. \eqno{(27)}$$
Thus, from Eq.(25) we can derive $n^{*}_{k}$ for the steady state:
$$n^{*}_{k} = \beta_{k}n^{*}_{k-1} = n^{*}_{m}\prod_{j=m+1}^{k} \beta_{j}$$
$$=\frac{\eta_{k-1}(k-1)^{\alpha}}{\frac{\mu_{\alpha}}{m}+\eta_{k}k^{\alpha}}
\cdots \frac{\eta_{m}m^{\alpha}}{\frac{\mu_{\alpha}}{m}+\eta_{m+1}(m+1)^{\alpha}}n^{*}_{m}$$
$$=\frac{\eta_{k-1}\cdots\eta_{m}\frac{\mu_{\alpha}}{m}\left[\frac{\Gamma(k)}{\Gamma(m)}\right]^{\alpha}}
{(\frac{\mu_{\alpha}}{m}+\eta_{k}k^{\alpha})\cdots(\frac{\mu_{\alpha}}{m}+\eta_{m}m^{\alpha})},\eqno{(28)}$$
where $k > m$ and in the last step we have used the gamma function 
$\Gamma (k) = (k-1)!$ and Eq.(27).

We can also rewrite the above in the same form
that was introduced by Krapivsky, Redner and Leyvraz\cite{Krapivsky00}
as follows:
$$n^{*}_{k} = \frac{\eta_{k-1}(k-1)^{\alpha}}{\frac{\mu_{\alpha}}{m}+\eta_{k}k^{\alpha}}
\cdots \frac{\eta_{m}m^{\alpha}}{\frac{\mu_{\alpha}}{m}+\eta_{m+1}(m+1)^{\alpha}}n^{*}_{m}$$
$$=\frac{\eta_{k}k^{\alpha}}{\frac{\mu_{\alpha}}{m}+\eta_{k}k^{\alpha}}
\cdots \frac{\eta_{m+1}(m+1)^{\alpha}}{\frac{\mu_{\alpha}}{m}+\eta_{m+1}(m+1)^{\alpha}}
\frac{\eta_{m}m^{\alpha}}{\eta_{k}k^{\alpha}}n^{*}_{m}$$
$$=\frac{\eta_{k}k^{\alpha}}{\frac{\mu_{\alpha}}{m}+\eta_{k}k^{\alpha}}
\cdots \frac{\eta_{m}m^{\alpha}}{\frac{\mu_{\alpha}}{m}+\eta_{m}m^{\alpha}} 
\frac{\frac{\mu_{\alpha}}{m}}{\eta_{k}k^{\alpha}}$$
$$=\frac{\frac{\mu_{\alpha}}{m}}{\eta_{k}k^{\alpha}}
\prod_{j=m}^{k}
\frac{1}{1+\frac{\mu_{\alpha}}{m\eta_{j}j^{\alpha}}}  \equiv P(k).  \eqno{(29)}$$
Substituting Eq.(29) into Eq.(15), the definition for $\mu_{\alpha}$,
we have the self-consistency condition for $\mu_{\alpha}$:
$$\mu_{\alpha} = \sum_{k=m}^{\infty}\eta_{k}k^{\alpha}n^{*}_{k} 
=\frac{\mu_{\alpha}}{m}
\sum_{k=m}^{\infty}\prod_{j=m}^{k}
\frac{1}{1+\frac{\mu_{\alpha}}{m\eta_{j}j^{\alpha}}}.  \eqno{(30)}$$
Then by expanding the right hand side with respect to $\mu_{\alpha}$
and subtracting some terms in both sides,
we obtain the simplified relation:
$$\mu_{\alpha} =\frac{1}{m}
\sum_{k=m+1}^{\infty}\prod_{j=m+1}^{k} 
\frac{1}{1+\frac{\mu_{\alpha}}{m\eta_{j}j^{\alpha}}}.  \eqno{(31)}$$
This equation corresponds to Eq.(6) of
Krapivsky, Redner and Leyvraz\cite{Krapivsky00}
and if we assume $\eta_{k} = 1$,
it coincides with Eq.(A.3) of Liu, Lai and Ye\cite{Liu02}
where the relation between $\alpha$ and $\mu_{\alpha}$
is demonstrated.

If we use the general notation $A_{k}$ for $m\eta_{k}k^{\alpha}$,
then we may rewrite Eq.(29) as  
$$P(k) \equiv n^{*}_{k} = \frac{\mu_{\alpha}}{A_{k}}
\prod_{j=m}^{k}
\frac{1}{1+\frac{\mu_{\alpha}}{A_{j}}},  \eqno{(32)}$$
which is equivalent to the Eq.(8) in Krapivsky, Redner and Leyvraz\cite{Krapivsky00}
if we take $m = 1$.
From the definition of the PA, it is obvious that
$A_{k}$ is nothing but the functional of PA.
Here we obtain
$$\mu_{\alpha} =A_{m}
\sum_{k=m+1}^{\infty}\prod_{j=m+1}^{k}
\frac{1}{1+\frac{\mu_{\alpha}}{A_{j}}}.  \eqno{(33)}$$

Eq.(29)[or Eq.(30)] include many cases.
If $\alpha = 1$ and $\eta_{k} = 1$, 
which is the case of Barab\'{a}si and Albert\cite{BarabasiAlbert99}
and produces
$\mu_{\alpha} = \mu_{1} = 2m$,
then 
$$P(k) \equiv n^{*}_{k} = \frac{2m(m+1)}{k(k+1)(k+2)} \propto k^{-3}.  \eqno{(34)}$$
If $\alpha = 1$ and $\eta_{k} \ne 1$, 
which is the case of Krapivsky, Redner and Leyvraz\cite{Krapivsky00}
and produces
$A_{\infty} = m\eta_{\infty}$,
then 
$$P(k) \equiv n^{*}_{k} \propto k^{-\gamma},  \eqno{(35)}$$
where $\gamma =1 + \mu_{1}/A_{\infty} = 1 + \mu_{1}/m\eta_{\infty}$
and $\mu_{1}$ can be calculated by using Eq.(30).

Thus, the heart of Simon's method is equivalent to that of 
Krapivsky, Redner and Leyvraz\cite{Krapivsky00}.
It provides a different result from that 
of Bornholdt and Ebel\cite{BornEbel01}.
This is due to the fact that both methods
adopt different assumptions to make the network systems,
although both use Simon's method.

\subsection{The Method of Dorogovtsev, Mendes and Samukhin}
Let us next consider how to solve the Eq.(18), once again.
This time we follow and generalize the method 
of Dorogovtsev, Mendes and Samukhin\cite{Dorogovtsev00}.
Recently, a similar approach has appeared\cite{Moore06}.

Going back to Eq.(17) and considering the steady state solution
we have
$$N_{k}(t) = \frac{m}{\mu_{\alpha}}[ -\eta_{k}k^{\alpha} N_{k}(t) 
+ \eta_{k-1}(k-1)^{\alpha} N_{k-1}(t)] + t\delta_{k,m}. \eqno{(36)}$$
Substituting $N_{k}(t) = n_{k}t$ into the above,
then we obtain
$$\left[\frac{\mu_{\alpha}}{m}+ \eta_{k}k^{\alpha} \right]n_{k} 
- \eta_{k-1}(k-1)^{\alpha} n_{k-1} 
= \frac{\mu_{\alpha}}{m}\delta_{k,m}. \eqno{(37)}$$
For the sake of simplicity for our purpose here, we take $\eta_{k} = 1$.
Then Eq.(37) becomes
$$\left[\frac{\mu_{\alpha}}{m}+ k^{\alpha} \right]n_{k} 
- (k-1)^{\alpha} n_{k-1} 
= \frac{\mu_{\alpha}}{m}\delta_{k,m}. \eqno{(38)}$$

Now, let us define a generating function:
$$\Phi_{m}(z) \equiv \sum_{k=m}^{\infty} n_{k}z^{k}.   \eqno{(39)}$$
Multiplying $z^{k}$ to Eq.(38) and summing up over $k$
from $m$ to $\infty$, we obtain
$$\left[\frac{\mu_{\alpha}}{m}+ 
(1-z)\left(z\frac{d}{dz}\right)^{\alpha} \right]\Phi_{m}(z) 
= \frac{\mu_{\alpha}}{m}z^{m}, \eqno{(40)}$$
where we have used the definition of the fractional derivative\cite{Nishimoto93,Hilfer00,Podlubny97}:
$$\left(z\frac{d}{d z}\right)^{\alpha} \Phi_{m}(z)
\equiv \sum_{k=m}^{\infty}k^{\alpha}n_{k}z^{k}.  \eqno{(41)}$$
When $\alpha = 1$, $\mu_{1}=2m$ and therefore, Eq.(40) turns out to be
$$\left[2 + (1-z)z\frac{d}{d z} \right]\Phi_{m}(z) = 2z^{m}. \eqno{(42)}$$

This is essentially equivalent to Eq.(7) in the paper of
Dorogovtsev, Mendes and Samukhin\cite{Dorogovtsev00}.

Let us show this.
Define as $\Phi(z) = \sum_{k=0}^{\infty}n_{k+m}z^{k} = n_{m} + n_{m+1}z + \cdots$,
which means that $k$ starts not from $m$ but from $0$. 
Therefore, we can rewrite $\Phi_m(z)$ as $\Phi_m(z) \equiv z^{m}\Phi(z)$.
From this, we find 
$$\frac{d}{d z} \left[\Phi_{m}(z)\right] =
\frac{d}{d z} \left[z^{m}\Phi(z)\right] 
= z^{m} \left(\frac{d}{d z} +\frac{m}{z}\right)\Phi(z).  \eqno{(43)}$$
Substituting Eq.(43) into Eq.(42), we obtain the following equation for $\Phi(z)$:
$$z(1-z)\frac{d}{d z} \Phi(z) + (m+2-mz) \Phi(z) = 2. \eqno{(44)}$$
By differentiation for both sides of Eq.(44) with respect to $z$,
we obtain 
$$z(1-z)\frac{d^{2}\Phi(z)}{dz^{2}} 
+ \left[m+3-(m+2)z\right] \frac{d\Phi(z)}{dz}-m\Phi(z) = 0. \eqno{(45)}$$
The solution of the above equation can be obtained by 
the Gauss' hypergeometric function:
$$_{2}F_{1}(a,b,c;x)
= \frac{\Gamma(c)}{\Gamma(a)\Gamma(b)}\sum_{n=0}^{\infty} 
\frac{\Gamma(a+n)\Gamma(b+n)}{\Gamma(c+n)n!}x^{n},\eqno{(46)}$$
which is the solution of the following ordinary differential equation:
$$x(1-x)\frac{d^{2}y(x)}{dx^{2}} + \left[c-(a+b+1)z\right]\frac{dy(x)}{dx}-ab y(x) = 0. \eqno{(47)}$$
Using the boundary condition that
$\Phi(0) =$ const and $\Phi'(0) = 0$
for Eq.(10), we get $\Phi(0) = 2/(m+2)$.
Using this and with replacement of $a = 1$, $b =m$, and $c = m+3$, 
$\Phi(z)$ can be obtained as
$$\Phi(z) = \frac{2}{m+2} \\ _{2}F_{1}(1,m,m+3;z)$$
$$= \sum_{k=0}^{\infty} 
\frac{2m(m+1)\Gamma(k +m)}{\Gamma(k+m+3)}x^{k}.  \eqno{(48)}$$
Therefore, comparing it with the definition of $\Phi(z)= \sum_{k=0}n_{k+m}z^{k}$,
we find
$$P(k) \equiv n_{k+m} = \frac{2m(m+1)\Gamma(k+m)}{\Gamma(k+m+3)}$$
$$= \frac{2m(m+1)}{(k+m)(k+m+1)(k+m+2)}.    \eqno{(49)}$$
Since in this expression $k$ starts from $k=0$,
if we change it to start from $k=m$
(i.e., we change the variable of $k$ from $k+m$ to $k$), then
we obtain
$$P(k) \equiv n_{k} = \frac{2m(m+1)}{k(k+1)(k+2)} \propto k^{-3}.    \eqno{(50)}$$
Hence, our result above is exactly equivalent to Eq.(34)
and coincides with that of Dorogovtsev, Mendes and Samukhin\cite{Dorogovtsev00}.
We can repeat the similar procedure for the integer $\alpha$ case.

Thus, we can understand that both methods of 
Krapivsky, Redner and Leyvraz\cite{Krapivsky00}
and 
Dorogovtsev, Mendes and Samukhin\cite{Dorogovtsev00}
give the identical result when $\alpha = 1$ 
and, therefore, they are essentially equivalent to each other.
However, this method is not directly applicable to
the case of fractional $\alpha$.
In this case we need a more sophisticated method
such as the {\it fractional calculus}\cite{Nishimoto93,Hilfer00,Podlubny97}.
This is a really challenging problem for physicists.
Some hints are discussed in Appendix A.


\section{The Solution of the Rate Equation for the Distribution Function}
In this section we are going to solve the rate equation of Eq.(20),
unifying both methods of 
Krapivsky, Redner and Leyvraz\cite{Krapivsky00}
and 
Dorogovtsev, Mendes and Samukhin\cite{Dorogovtsev00}. 

\subsection{The Discrete Rate Equation}
Let us first apply the method  
of Krapivsky, Redner and Leyvraz\cite{Krapivsky00}. 
Now, we want to solve the discrete rate equation:
$$N_{k}(i,t+1)  = N_{k}(i,t) - \frac{m\eta_{k}k^{\alpha}}{\mu_{\alpha} t}N_{k}(i,t) $$
$$+ \frac{m\eta_{k-1}(k-1)^{\alpha}}{\mu_{\alpha} t}N_{k-1}(i,t).\eqno{(51)}$$
Let us define the average connectivity $\overline{k}(i,t)$
by
$$\overline{k}(i,t) \equiv \sum_{k=m}^{\infty}kN_{k}(i,t). \eqno{(52)}$$
Multiplying Eq.(20) by $k$ and summing up over $k$,
and using this definition
we obtain
$$\overline{k}(i,t+1) = \overline{k}(i,t)
+ \frac{m}{\mu_{\alpha} t}\overline{\eta_{k}k^{\alpha}}(i,t),\eqno{(53)}$$
where $\overline{\eta_{k}k^{\alpha}}(i,t) \equiv \sum_{k=m}^{\infty}\eta_{k}k^{\alpha}N_{k}(i,t)$. 
If we set $\eta_{k} = 1$, then it becomes
$$\overline{k}(i,t+1) = \overline{k}(i,t)
+ \frac{m}{\mu_{\alpha} t}\overline{k^{\alpha}}(i,t),\eqno{(54)}$$
where  
$\overline{k^{\alpha}}(i,t) \equiv \sum_{k=m}^{\infty}k^{\alpha}N_{k}(i,t)$.
Hence, if we can further impose $\alpha =1$,
then we obtain 
$$\overline{k}(i,t+1) = \overline{k}(i,t) + \frac{1}{2 t}\overline{k}(i,t),  \eqno{(55)}$$
where we have used $\mu_{1} = 2m$.

Eq.(54) can be regarded as a discrete version of the continuous time equation 
such as Eq.(2) with the PA $\Pi_{i}$ of fractional exponent $\alpha$ :
$$\frac{dk_{i}}{dt} = m\Pi_{i} = \frac{m k_{i}^{\alpha}}{\sum_{j=1}^{N-1}k_{j}}
=  \frac{m k_{i}^{\alpha}}{\mu_{\alpha}t}.  \eqno{(56)}$$
In general, since $\overline{k^{\alpha}}(i,t)$ is not equal to $\overline{k}_{i}^{\alpha}
= [\overline{k}(i,t)]^{\alpha}$,
the discretization of Eq.(56) is not equivalent to Eq.(54).
However, for $\alpha = 1$, Eq. (56) becomes identical to Eq.(2)
and therefore, the discretization of Eq.(2) becomes equivalent to Eq.(55).
Eq.(55) can be solved directly as follows:
From Eq.(55) we find
$$\overline{k}(i,t+1) = \left(1+\frac{1}{2t}\right)\overline{k}(i,t)$$
$$= \left(1+\frac{1}{2t}\right)\cdots \left(1+\frac{1}{2(t-t_{i})}\right)
\overline{k}(i,t_{i}). \eqno{(57)}$$
Taking logarithm of both sides, we obtain
$$\ln \overline{k}(i,t) 
= \sum_{j=t_{i}}^{t-1}\ln \left(1+\frac{1}{2j}\right)
+\ln\overline{k}(i,t_{i})$$
$$\approx \frac{1}{2}\sum_{j=t_{i}}^{t-1}\frac{1}{j} +\ln\overline{k}(i,t_{i})
\approx \frac{1}{2}\ln\left(\frac{t}{t_{i}}\right) +\ln\overline{k}(i,t_{i}), \eqno{(58)}$$
where we have assumed that $t$ is sufficiently large such that $t \gg t_{i}$.
Hence, we approximately obtain
$$\overline{k}(i,t) = m\left(\frac{t}{t_{i}}\right)^{1/2},  \eqno{(59)}$$
where we have used $\overline{k}(i,t_{i}) = m$.
This exactly coincides with the result of
time development of the seminal work
of Barab\'{a}si and Albert's continuous model\cite{BarabasiAlbert99}.

To go further to the case of fractional $\alpha$, the discrete model is not 
so convenient since Eq.(53) is not simple equation of $\overline{k}(i,t)$.
Thus, for this purpose, we have to use the continuous rate equation.

\subsection{The Fractional Differential Rate Equation}
Let us now take a continuous time limit,
by which we obtain the following continuous differential equation:
$$\frac{\mu_{\alpha}}{m}t \frac{\partial}{\partial t}N_{k}(i,t)  
= - \eta_{k}k^{\alpha}N_{k}(i,t) 
+ \eta_{k-1}(k-1)^{\alpha} N_{k-1}(i,t),\eqno{(60)}$$
where $N_{k}(i,t)$ is the connectivity distribution function of node $i$ at time $t$.
For our purpose here, we will set $\eta_{k} = \eta =$ constant,
for the sake of simplicity.

Let us define the $Z$-transformation of the connectivity distribution function $N_{k}(i,t)$
by
$$\Phi_{m}(i, t; z) \equiv \sum_{k=m}^{\infty}N_{k}(i,t)z^{k}
\equiv z^{m}\Phi(i,t;z), \eqno{(61)}$$
where $\Phi(i,t;z) \equiv \sum_{k=0}^{\infty}N_{k+m}(i,t)z^{k}$.
Multiplying both sides of Eq.(60) by $z^{k}$ and summing up over $k$
and using the fractional derivative that was introduced in the previous section,
we obtain the following fractional differential equation:
$$\frac{1}{\beta_{\alpha}}t \frac{\partial}{\partial t}\Phi_{m}(i,t;z)  
+ (1-z) \left(z\frac{\partial}{\partial z}\right)^{\alpha}\Phi_{m}(i,t;z)= 0,\eqno{(62)}$$
where $\beta_{\alpha} \equiv \frac{m\eta}{\mu_{\alpha}}$.
This corresponds to Eq.(40) in Sec.IV
and also it corresponds to Eq.(14) 
in 
Krapivsky, Redner and Leyvraz\cite{Krapivsky00}.

\subsection{The Solution of the Rate Equation with Linear Preferential Attachment}
When $\alpha$ is fractional we meet the same difficulty that
has been discussed in the end of the previous section.
Therefore, let us restrict ourselves to consider the simplest case of linear PA with
$\alpha = 1$.
In this case, $\mu_{1} = 2m$ such that $\beta_{1} = \frac{1}{2\eta}$
and therefore, Eq.(62) becomes
$$\frac{1}{\beta_{1}}t \frac{\partial}{\partial t}\Phi_{m}(i,t;z)  
+ (1-z) z\frac{\partial}{\partial z}\Phi_{m}(i,t;z)= 0.   \eqno{(63)}$$
If we take $\eta = 1$ this corresponds to Eq.(42)
and if $\eta > 1$ then it corresponds to Eq.(14) in 
Dorogovtsev, Mendes and Samukhin\cite{Dorogovtsev00}.

Let us solve Eq.(63).
Using the standard method for solving linear partial differential
equation, it yields the following differential equations
for the characteristic curve:

$$\frac{dz}{z(1-z)}=\beta_{1}\frac{dt}{t} = \frac{d\Phi_{m}}{0},      \eqno{(64)}$$
where $0$ in the denominator in the last equation means that 
$d\Phi_{m} = 0$
such that
$\Phi_{m}(i,i;z) = c_{1} =$ constant.
Since by definition $N_{k}(i,i) = \delta_{k,m}$,
we have the initial condition
$$\Phi_{m}(i,i;z) = c_{1} = z_{i}^{m},                    \eqno{(65)}$$
where $z_{i}$ means the initial value of $z$ at time $t_{i} = i$.
Solving the first relation yields
$$\frac{z}{1-z} = c_{2}\left(\frac{t}{t_{i}}\right)^{\beta_{1}},   \eqno{(66)}$$
which then gives
$$\frac{z_{i}}{1-z_{i}} = c_{2},   \eqno{(67)}$$
Substituting Eq.(67) into Eq.(65), we obtain the relation:
$$c_{1} = \left(\frac{c_{2}}{1+c_{2}} \right)^{m},   \eqno{(68)}$$
which is the characteristic curve that we seek for Eq.(64).
Since the solution moves as a point along this curve from 
the initial time $t=t_{i}$,
the solution of Eq.(63) is given by
$$\Phi_{m}(i,t;z) = c_{1}(i,t; z) 
= \left[\frac{c_{2}(i,t; z)}{1+c_{2}(i,t; z)} \right]^{m}.   \eqno{(69)}$$
Now, solving Eq.(66) for $c_{2}$ and substituting it into Eq.(69),
we obtain
$$\Phi_{m}(i,t;z) 
= \left[\frac{z\left(t_{i}/t\right)^{\beta_{1}}}
{1-z\left(1-\left(t_{i}/t\right)^{\beta_{1}}\right)} \right]^{m}.   \eqno{(70)}$$
Adjusting with our definition of Eq.(61),
we finally obtain the solution of Eq.(62)
as
$$\Phi(i,t;z) 
= \left[\frac{\left(t_{i}/t\right)^{\beta_{1}}}
{1-z\left(1-\left(t_{i}/t\right)^{\beta_{1}}\right)} \right]^{m}.   \eqno{(71)}$$
Now expanding the denominator of Eq.(71) using the formula:
$$\frac{1}{(1-z)^{\gamma}} 
= \sum_{k=0}^{\infty} \frac{\Gamma(k+\gamma)}{\Gamma(\gamma)k!}z^{k},  \eqno{(72)}$$
we obtain
$$\Phi(i,t;z) 
= \sum_{k=0}^{\infty}\frac{\Gamma(k+m)}{\Gamma(m)k!}
\left(\frac{t_{i}}{t}\right)^{m\beta_{1}}
\left[1-\left(\frac{t_{i}}{t}\right)^{\beta_{1}}\right]^{k}z^{k}.   \eqno{(73)}$$
Comparing this with Eq.(61), 
we end up with the following distribution function:
$$N_{k+m}(i, t) =  \frac{\Gamma(k+m)}{\Gamma(m)k!}
\left(\frac{t_{i}}{t}\right)^{m\beta_{1}}
\left[1-\left(\frac{t_{i}}{t}\right)^{\beta_{1}}\right]^{k}, \eqno{(74)}$$
where $\sum_{k=0}^{\infty}N_{k+m}(i, t) = 1$ 
is obviously satisfied.
The Eq.(74) corresponds to Eq.(16) in Krapivsky, Redner and Leyvraz\cite{Krapivsky00}
and Eq.(15) in
Dorogovtsev, Mendes and Samukhin\cite{Dorogovtsev00}.

\subsection{Time Development of the Average Connectivity}
Applying  Eq.(74) together with the following formula:
$$\frac{\gamma z}{(1-z)^{\gamma}} 
= \sum_{k=0}^{\infty} k\frac{\Gamma(k+\gamma)}{\Gamma(\gamma)k!}z^{k},  \eqno{(75)}$$
the time development of the average connectivity of node $i$ 
is calculated as
$$\overline{k}(i,t)  = \sum_{k=0}^{\infty} k N_{k+m}(i, t)
= m\left(\frac{t}{t_{i}}\right)^{\beta_{1}},  \eqno{(76)}$$
which corresponds to Eq.(59) for the case of the discrete rate equation.
If there is no effect of the preference so that $\eta = 1$, then $\beta_{1} = 1/2$;
In this case, Eq.(76) recovers the same time development of the seminal work
of Barab\'{a}si and Albert\cite{BarabasiAlbert99} once again.
And also it corresponds to Eq.(16) in 
Dorogovtsev, Mendes and Samukhin\cite{Dorogovtsev00}, respectively.

Notify Eq.(35) for the scale-free exponent of $\gamma_{1}$ for $\alpha = 1$,
where $\gamma_{1} = 1 + \frac{\mu_{1}}{m\eta_{\infty}} = 1 + \frac{2}{\eta_{\infty}}$. 
If we set $\eta = \eta_{\infty}$ then
from the definition of $\beta_{1} = \eta/2$
we find the relationship between $\gamma_{1}$ and $\beta_{1}$:
$$\beta_{1} = \frac{1}{\gamma_{1}-1}.   \eqno{(77)}$$
This recovers Eq.(17) in Dorogovtsev, Mendes and Samukhin\cite{Dorogovtsev00}.
As pointed out by them, this relation is {\it exact} for any finite preference of $\eta$
as long as $\alpha = 1$.
However, we do not know yet whether or not this relation holds true
even for the fractional $\alpha$
such as $\beta_{\alpha} = 1/(\gamma_{\alpha}-1)$.


\section{Conclusions}
In conclusion, we have studied an analytical method 
of how to obtain the connectivity distribution functions 
for the various growing networks with PA of fractional power.
First, in order to unify the growing network models
in a general point of view, 
we have discussed the various types of PA [See Eq.(8)].
Second, we have presented a general version of the rate equations
for the growing networks with PA of fractional power [See Eq.(16)].
Here we have presented the rate equations 
for the connectivity distribution functions for the entire network 
and for an arbitrary node, respectively.
Third, we have presented the way of solving the rate equations
for connectivity distribution functions for the entire network
from the point of view of Simon's method\cite{Simon55}.
We have shown that our method unifies both methods of
Krapivsky, Redner and Leyvraz\cite{Krapivsky00}
and 
Dorogovtsev, Mendes and Samukhin\cite{Dorogovtsev00}
into the same mathematical framework.
We have introduced an idea of fractional calculus that 
the rate equation for the connectivity distribution function with
PA of fractional power
becomes the fractional differential equation.
We have presented a scheme to solve the problem in the appendix,
where we have introduced a new type of transidental function
that we call the {\it upsilon function}, $\Upsilon(a,b,c;z)$[See A.13].
Fourth, we have presented the way of solving the rate equations
for connectivity distribution functions for the $i$th node,
using the generalize method.

Thus, we have unified some of the previously known methods and results of connectivity
distribution functions for such growing networks into a single analytical theory
using fractional calculus
in the spirit of the seminal Simon's method of word problem\cite{Simon55}.
In this context, we have emphasized that in spirit,
Simon's method is identical not only to the method of
Bornholdt and Ebel\cite{BornEbel01}
but also to both methods of 
Krapivsky, Redner and Leyvraz\cite{Krapivsky00}
and 
Dorogovtsev, Mendes and Samukhin\cite{Dorogovtsev00}.
However, the former provides a different scale-free exponent from the latter,
since the defining conditions for the growing networks are distinct to each other.
We have been able to establish a new method of fractional calculus to the 
problem of growing network models with PA of fractional power of $\alpha$.
This type of the fractional differential equation [See Eq.(40)] seems to be 
not yet known  in this field of growing network models,
which is different from the one that is known
as the fractional diffusion equations etc in the fractional calculus
\cite{Nishimoto93,Hilfer00,Podlubny97}.
Therefore, we believe that our approach may shed a new light on
solving such a novel type of fractional differential equations in its own right.

\acknowledgments
We would like to thank 
Shu-ichi Kinoshita and Dr. Jun Hidaka for sending us many relevant papers
and Dr. Katsuyuki Nishimoto and Dr. Kazuhiko Aomoto
for very valuable suggestions and comments.
K. I. would like to give special thanks to Kazuko Iguchi for her continuous
financial support and encouragement.

\appendix
\section{Fractional Calculus}
For the noninteger case of $\alpha$, 
we need a different way of solving the rate equation.
We have to invent a new method.
This case seems to require the method of {\it fractional calculus}
\cite{Nishimoto93,Hilfer00,Podlubny97}.

Let us see this point.
Go back to Eq.(40), once again. 
$$\left[\frac{\mu_{\alpha}}{m}+ 
(1-z)\left(z\frac{d}{d z}\right)^{\alpha} \right]\Phi_{m}(z) 
= \frac{\mu_{\alpha}}{m}z^{m}.                        \eqno{(40)}$$
This type of differential equation has been called the {\it fractional differential equation}
when the degree of the derivative is fractional,
which is called {\it fractional derivative}.
Recently, there have appeared many applications of the concept to
various areas of physics\cite{Nishimoto93,Hilfer00,Podlubny97}.

From this context, our problem is one of such application to physics as well.
How to solve the equation such as Eq.(40) with fractional or irrational $\alpha$
would be a mathematical challenge to physicists.

Using the properties of fractional differentials that
$(\frac{d}{d z})^{\alpha} =(\frac{d}{d z})(\frac{d}{d z})^{\alpha-1} $,
let us rewrite Eq.(40) as
$$\left[\frac{\mu_{\alpha}}{m}+ 
z(1-z)\left(\frac{d}{d z}\right)
\left(z\frac{d}{d z}\right)^{\alpha-1} \right]\Phi_{m}(z) 
= \frac{\mu_{\alpha}}{m}z^{m}, \eqno{(A.1)}$$
which corresponds to Eq.(42).
Using the relation of Eq.(43),
Eq.(A.1) becomes
$$\left[\frac{\mu_{\alpha}}{m}+ 
z(1-z)\left(\frac{d}{d z} + \frac{m}{z}\right)
\left(z\frac{d}{d z} + m\right)^{\alpha-1} \right]\Phi(z) 
= \frac{\mu_{\alpha}}{m},               \eqno{(A.2)}$$
which corresponds to Eq.(44).
Operating both sides of Eq.(A.2) by $z\frac{\partial}{\partial z}$,
it turns out to be the following fractional differential equation:
$$\frac{\mu_{\alpha}}{m}z\frac{d}{d z}\Phi(z)
+
(1-z)\left(z\frac{d}{d z} +m\right)^{\alpha+1}\Phi(z)$$
$$+ \left[-m+(m-1)z\right]\left(z\frac{d}{d z} +m\right)^{\alpha}\Phi(z) 
= 0,                       \eqno{(A.3)}$$
which corresponds to Eq.(45).
Comparing this with Eq.(47),
we now easily see that if $\alpha = 1$, then Eq.(A.3) turns out to be Eq.(48).
Thus, the problem is mapped to how to solve the fractional differential equation
such as Eq.(40) or Eq.(A.2) when $\alpha$ is fractional.

Let us denote as $z = e^{u}$.
Then, Eq.(40) and Eq.(A.2) can be written as
$$\left[\frac{\mu_{\alpha}}{m}+ 
(1-e^{u})\left(\frac{d}{d u}\right)^{\alpha} \right]\Phi_{m}(u) 
= \frac{\mu_{\alpha}}{m}e^{mu}, \eqno{(A.4)}$$
and
$$\left[\frac{\mu_{\alpha}}{m}+ 
(1-e^{u})\left(\frac{d}{d u} +m\right)^{\alpha} \right]\Phi(u) 
= \frac{\mu_{\alpha}}{m}, \eqno{(A.5)}$$
respectively.
Similarly, Eq.(A.3) is transformed into the following:
$$\frac{\mu_{\alpha}}{m}\frac{d}{d u}\Phi(u)
+
(1-e^{u})\left(\frac{d}{d u} +m\right)^{\alpha+1}\Phi(u)$$
$$+\left[-m+(m-1)e^{u}\right]\left(\frac{d}{d u} +m\right)^{\alpha}\Phi(u) 
= 0,                                       \eqno{(A.6)}$$

Let us now define the Laplace transformation:
$${\cal L}[\Phi(u)] \equiv \Psi(s) = \int_{0}^{\infty} \Phi(u)e^{-s u} du.\eqno{(A.7)}$$
The definition of fractional derivative requires the following operation:
$${\cal L}\left[\left(\frac{d}{d u}+m\right)^{\alpha}\Phi(u)\right] 
= (s+m)^{\alpha} \Psi(s).                          \eqno{(A.8)}$$
The Laplace transforms of other terms are given as follows:
$${\cal L}\left[e^{u}\Phi(u)\right] =\Psi(s-1).          \eqno{(A.9)}$$ 
$${\cal L}\left[e^{u}\left(\frac{d}{d u}+m\right)^{\alpha}\Phi(u)\right] 
=(s+m-1)^{\alpha}\Psi(s-1).                            \eqno{(A.10)}$$ 
The proof of these relations is straightforward.
Using these, the Laplace transform of Eq.(A.6) becomes
$$\frac{\mu_{\alpha}}{m}s\Psi(s) +(s +m)^{\alpha+1}\Psi(s)
-(s +m -1)^{\alpha+1}\Psi(s-1)$$
$$- m(s +m)^{\alpha}\Psi(s) 
+(m-1)(s +m -1)^{\alpha}\Phi(s-1) = 0,                    \eqno{(A.11)}$$
which yields a recursion relation:
$$\Psi(s)
= \frac{(s +m -1)^{\alpha}}
{\frac{\mu_{\alpha}}{m} +(s +m)^{\alpha}}\Psi(s-1).       \eqno{(A.12)}$$

Please compare this with Eq.(25) in Sec.IV.
Since we take $\eta_{k} = 1$ for our purpose here, 
it is clear that
Eq.(A.12) in the language of the (continuous) Laplace transformation corresponds to 
Eq.(25) in the language of the (discrete) $Z$-transformation,
where the continuous variable $s+m$ in Eq.(A.12) 
corresponds to the discrete variable $k$ in Eq.(25).

Looking at Eq.(A.12), if we are able to Laplace transform  $\Psi(s)$
into the original function $\Phi(z)$, then the problem can be solved.
However, it is not so simple since Eq.(A.12) is a recursion relation for $\Psi(s)$
and the variable $s$ is a {\it continuous} one.  
Therefore, we cannot simply represent $\Psi(s)$ in terms of $s$ 
such as $n^{*}_{k}$ in the case of Eq.(25).
Rather, it should be regarded as a functional relation.
For example, it is supposed to be like a functional relation such as 
$\Gamma(z+1) = z\Gamma(z)$,
which defines the Gamma-function $\Gamma(z)$,
where if $z = n$, integer, then $\Gamma(n+1) = n!$.

From this understanding,
we would like to regard the relation (A.12) as a defining relation
for a new (unknown) type of function.
Let us define the following function $\Upsilon_{\alpha}(a,b,c;s)$, upsilon function of $s$,
such that
$$\Upsilon_{\alpha}(a,b,c;s) = \frac{b(s+c-1)^{\alpha}}
{a + b (s+c)^{\alpha}}\Upsilon_{\alpha}(a,b,c;s-1),      \eqno{(A.13)}$$
where $a$, $b$, $c$ are real numbers.
From this, as suggested by Aomoto\cite{Aomoto07},
we find that the upsilon function can be represented
in terms of an infinite product:
$$\frac{1}{\Upsilon_{\alpha}(a,b,c;s)} \equiv \frac{1}{b(s+c)^{\alpha}}
\prod_{n=1}^{\infty}\frac{1}{1+\frac{a}{b}\frac{1}{(s+c+n)^{\alpha}}}.   \eqno{(A.14)}$$
The convergence of the upsilon function can be dominated by
the function of an infinite product in the denominator.
If we denote it by
$$f_{\alpha}(s;x) \equiv \prod_{n=1}^{\infty}\left( 1 + \frac{x}{(s + n)^{\alpha}}\right), \eqno{(A.15)}$$
then from Eq.(A.14) the upsilon function can be obtained as
$$\Upsilon_{\alpha}(a,b,c;s) = b (s+c)^{\alpha}f_{\alpha}(s+c;a/b).      \eqno{(A.16)}$$
The analytical properties and convergence of this function $f_{\alpha}(s;x)$ may be
obtained by the standard argument of entire functions\cite{Boas54}.

At this moment, we do not know much about the mathematical properties 
of the function $\Upsilon_{\alpha}(a,b,c;s)$\cite{Aomoto07}.
However, if we define this function, then we can formally represent
the $Z$-transformed function $\Phi(z)$ in terms of the upsilon function
as follows.
Adjusting Eq.(A.12) with our definition of (A.13),
our $\Psi(s)$ can be written in terms of $\Upsilon_{\alpha}(a,b,c;s)$ as
$$\Psi(s) \equiv \Upsilon_{\alpha}\left(\frac{\mu_{\alpha}}{m},1,m;s\right).  \eqno{(A.17)}$$
Now, we invert $\Psi(s)$ by the inverse Laplace transformation:
$$\Phi(u) = \frac{1}{2\pi i}\int_{-\infty+c i}^{\infty+c i}\Psi(s)e^{su}ds$$
$$= \frac{1}{2\pi i}\int_{-\infty+c i}^{\infty+c i}
\Upsilon_{\alpha}\left(\frac{\mu_{\alpha}}{m},1,m;s\right)e^{su}ds, \eqno{(A.18)}$$
where $c$ is a constant.
Replacing as $z = e^{u}$, we finally obtain the following formula for $\Phi(z)$:
$$\Phi(z) = \frac{1}{2\pi i}\int_{-\infty+c i}^{\infty+c i}
\Upsilon_{\alpha}\left(\frac{\mu_{\alpha}}{m},1,m;s\right)z^{s}ds. \eqno{(A.19)}$$
This is supposed to be the solution of the fractional differential equation of Eq.(A.3).
 
Next, if we can expand $\Phi(z)$ in terms of $z$
such as 
$$\Phi(z) = \sum_{k=0}^{\infty} \overline{\Upsilon}_{\alpha}
\left(\frac{\mu_{\alpha}}{m},1,m;k\right)z^{k},  \eqno{(A.20)}$$
where $\overline{\Upsilon}_{\alpha}
\left(\frac{\mu_{\alpha}}{m},1,m;k\right)$ denote the expansion coefficients, 
then comparing the definition of the $Z$-transformation:
$\Phi(z) = \sum_{k=0}^{\infty} n_{k+m} z^{k}$,
we obtain 
$$n_{k+m} = \overline{\Upsilon}_{\alpha}
\left(\frac{\mu_{\alpha}}{m},1,m;k\right).    \eqno{(A.21)}$$
Hence, our problem is solved.



\begin{references}

\bibitem[*]{byline1} e-mail: hyamada@uranus.dti.ne.jp.

\bibitem{ErdosRenyi} 
P. Erd\"{o}s and A. R\'{e}nyi, 
Publ. Math. {\bf 6}, 290 (1959);
Publications of the Mathematical Institute of the Hungarian Academy of Sciences {\bf 5}, 17 (1960);
Acta Mathematica Acadamiae Scientiorum Hungaricae {\bf 12} 261 (1961).

\bibitem{Strogatz01}
S. H. Strogatz,
Nature {\bf 410}, 268-276 (2001).

\bibitem{Barabasi02}
A.-L. Barab\'{a}si, 
{\it Linked}, 
(Penguin books, London, 2002)

\bibitem{AlbertBarabasi02}
R. Albert and A.-L. Barab\'{a}si, 
Rev. Mod. Phys. {\bf 74}, 47-97 (2002).

\bibitem{BaraBona03}
A.-L. Barab\'{a}si and E. Bonabeau, 
Sci. Am. {\bf 288}(May), 60-69 (2003).

\bibitem{Newman03}
M. E. J. Newman, 
SIAM Rev. {\bf 45}, 167-256 (2003).    

\bibitem{DorogovMendes03}
S. N. Dorogovtsev and J. F. F. Mendes, 
{\it Evolution of Networks: From Biological Nets to the Internet and WWW\/}, 
(Oxford University, New York, 2003); 
Adv. Phys. {\bf 51}, 1079 (2002);
cond-mat/0204102.

\bibitem{Boccaletti03}
S. Boccaletti, V. Latora, Y. Moreno, M. Chavez, and D.-U. Hwang,
Physics Reports {\bf 424}, 175-308 (2006). 


\bibitem{BarabasiAlbert99}
A.-L. Barab\'{a}si and R. Albert, 
Science {\bf 286}, 509-512 (1999).

\bibitem{DorogovMendes00}
S. N. Dorogovtsev and J. F. F. Mendes, 
Europhys. Lett. {\bf 52}, 33-39 (2000).
 
\bibitem{Krapivsky00} 
P. L. Krapivsky, S. Redner, and F. Leyvraz, 
Phys. Rev. Lett. {\bf 85}, 4629-4632 (2000).
Also see
P. L. Krapivsky and S. Redner,
Phys. Rev. E {\bf 63}, 066123 (2001).
P. L. Krapivsky, G. J. Rodgers, and S. Redner,
Phys. Rev. Lett. {\bf 86}, 5401-5404 (2001).

\bibitem{Dorogovtsev00}
S. N. Dorogovtsev, J. F. F. Mendes, and A. N. Samukhin, 
Phys. Rev. Lett. {\bf 85}, 4633-4636 (2000).
Also see
S. N. Dorogovtsev, J. F. F. Mendes, 
Phys. Rev. E {\bf 63}, 056125-1-19 (2000).
S. N. Dorogovtsev, A. V. Goltsev, J. F. F. Mendes and A. N. Samukhin, 
Phys. Rev. E {\bf 68}, 046109-1-10 (2003).

\bibitem{BarabasiAlbert00}
R. Albert and A.-L. Barab\'{a}si, 
Phys. Rev. Lett. {\bf 85}, 5234-5237 (2000).

\bibitem{Cohen00}
R. Cohen, K. Erez, D. ben-Avraham, and S. Havlin, 
Phys. Rev. Lett. {\bf 85}, 4626-4628 (2000).

\bibitem{Callaway01}
D. S. Callaway, J. E. Hopcroft, J. M. Kleinberg, M. E. J. Newman, 
and S. H. Strogatz, 
Phys. Rev. E {\bf 64}, 041902-1-7 (2001).

\bibitem{BornEbel01}
S. Bornholdt and H. Ebel, 
Phys. Rev. E {\bf 64}, 035104(R)-1-4 (2001).

\bibitem{BianconiBara01} 
G. Bianconi and A.-L. Barab\'{a}si, 
Phys. Rev. Lett. {\bf 86}, 5632-5635 (2001).

\bibitem{Goh01}
K.-I. Goh, B. Kahng, and D. Kim, 
Phys. Rev. Lett. {\bf 87}, 278701-1-4 (2001).

\bibitem{Liu02}
Z. Liu, Y.-C. Lai, and N. Ye, 
Phys. Rev. E {\bf 66}, 036112-1-7 (2002).
Also see
R. N. Onody and P. A. de Castro, 
Physica A {\bf 336}, 491-502 (2004).


\bibitem{Milo02}
R. Milo, S. Shen-Orr, S. Itzkovitz, N. Kashtan, D. Chklovskii and U. Alon, 
Science {\bf 298}, 824-827 (2002).

\bibitem{Przytycka04}
T. M. Przytycka and Y.-K. Yu, 
Compt. Biol. Chem. {\bf 28}, 257-264 (2004).

\bibitem{Song05}
C. Song, S. Havlin, and H. A. Makse, 
Nature {\bf 433}, 392-395 (2005).

\bibitem{Catanzaro05}
M. Catanzaro, M. Bogu\~{n}\'{a}, and R. Pastor-Satorras, 
Phys. Rev. E {\bf 71}, 027103(4 pages) (2005). 

\bibitem{White05}
D. R. White, N. Kej\v{z}ar, C. Tsallis, D. Farmer and S. White, 
Phys. Rev. E {\bf 73}, 016119(8 pages) (2006). 

\bibitem{TakemotoOosawa05}
K. Takemoto and C. Oosawa, 
Phys. Rev. {\bf E 72}, 046116 (2005).

\bibitem{Moore06}
C. Moore, G. Ghoshal, and M. E. J. Newman, 
Phys. Rev. E {\bf 74}, 036121(9 pages) (2006). 

\bibitem{Bu07}
S. Bu, B.-H. Wang, and T. Zhou, 
Physica A {\bf 374}, 864-868 (2007).




\bibitem{Simon55}
H. A. Simon, 
Biometrika {\bf 42}, 425-440 (1955).

\bibitem{Footnote1}
A similar critique has been presented in Ref.
\cite{AlbertBarabasi02}.

\bibitem{Footnote2}
Here, to adjust with the results of Lie, Lai and Ye(Ref.\cite{Liu02}),
we take the situation that the node $i$ is born with $m$ links
such that $k_{i}(t_{i}) = m$.
However, if we want to adjust with the results 
of Krapivsky, Redner and Leyvraz(Ref.\cite{Krapivsky00})
then we can just adopt $\delta_{k,1}$, since
in their model each node starts with $k=1$.
On the other hand,
if we want to adjust with the results of 
Dorogovtsev, Mendes and Samukhin\cite{Dorogovtsev00},
then we adopt $\delta_{k,0}$, since
in their model each node starts with $k=0$.
Thus, the value of $m$ in the $\delta$-function
depends on the model that we adopt.

\bibitem{Nishimoto93}
K. Nishimoto,
{\it Kummer's Twenty-Four Functions and $N$-fractional Calculus\/}, in
Nonlinear Analysis, Theory, Methods \& Applications
{\bf 30}, 1271-1282 (1997);
J. Fractional Calculus {\bf 15}, 
67-72, 
73-82, 
83-89 (1999)
and references therein.


\bibitem{Hilfer00} 
R. Hilfer ed., 
{\it Applications of Fractional Calculus in Physics\/},
(World Scientific, Singapore, 2000).
I. M. Sokolov, J. Klafter, and A. Blumen,
{\it Fractional Calculus\/},
Phys. Today {\bf 55}(November), 48-52 (2002).

\bibitem{Podlubny97} 
I. Podlubny, 
{\it The Laplace Transform Method for Linear Differential 
Equations of the Fractional Order\/},
{\it Fractional Differential Equations}
(Academic, New York, 1999) or
funct-an/9710005.

\bibitem{Aomoto07}
K. Aomoto, private communications (2 March 2007).
This function $f_{\alpha}(s,x)$ has the following properties:
For $\alpha > 1$, the order $\rho$, the convergence order $\rho_{1}$
and the genus $p$ of the function are given as
$\rho = \rho_{1} = 1/\alpha < 1$,
and $p = [1/\alpha] = 0$, respectively;
for $0 < \alpha < 1$, $\rho = \rho_{1} = 1/\alpha > 1$, 
and $p = [1/\alpha] > 1$;
and for $\alpha = 1$, the function is the $\Gamma$-function type.
Here $[ \ ]$ is the Gauss symbol that takes the largest integer part.

\bibitem{Boas54} 
R. P. Boas, Jr.,
{\it Entire Functions\/},
(Academic, New York, 1954).


\end{references}
\end{document}